\documentclass[preprint,prd,aps,showpacs,showkeys,nofootinbib]{revtex4}
\usepackage{amsmath}
\usepackage{graphicx}
\usepackage{diagbox}
\usepackage{slashed}
\usepackage{float}
\usepackage{color}
\begin{document}
\title{ The partial waves of $B^{*}_{2}(5747)$ and their contributions in strong decays}
\author{Ting-Ting Liu$^{1,2}$\footnote{liutt155@163.com},
Su-Yan Pei$^{1,2}$,
Wei Li$^{1,2}$,
Meng Han$^{1,2}$,
Guo-Li Wang$^{1,2}$\footnote{wgl@hbu.edu.cn, corresponding author}}
\affiliation{$^1$ Department of Physics and Technology, Hebei University, Baoding 071002, China\\
$^2$ Key Laboratory of High-Precision Computation and Application of Quantum Field Theory of Hebei Province, Baoding, China}
\begin{abstract}
The relativistic Bethe$-$Salpeter method is adopted to study the OZI allowed strong decays of the $2^+$ state $B^{*}_{2}(5747)^{0}$, with emphasis on the relativistic corrections. We first study the partial waves in the wave functions used and find that there are $P$, $D$ and $F$ partial waves in $B^{*}_{2}(5747)^{0}$ meson, with ratios of $P/D/F=1:0.421:0.051$. We also find $S/P/D=1:0.354:0.046$ for $B^*$, and $S/P=1:0.343$ for $B$ meson. The large components of the $D$ wave in $B^{*}_{2}(5747)^{0}$ and the $P$ wave in $B^{(*)}$ mean that large relativistic effects exist in these states. Second, we calculate the strong decays, and the total decay width $\Gamma(B^{*}_{2}(5747)^{0})=25.9$ MeV and the branching fraction  $\Gamma(B^{*}_{2}(5747)^{0} \to B^{*}\pi)$ / $\Gamma(B^{*}_{2}(5747)^{0} \to B\pi)=0.96$ obtained are consistent with experimental data. Third, we study the contributions of different partial waves in the initial and final wave functions, and find that the relativistic effects are about $15\%$ and $11\%$ for $B^{*}_{2}(5747)^{0} \to B\pi$ and  $B^{*}_{2}(5747)^{0} \to B^{*}\pi$, respectively, which are much smaller than our expected effects, showing that the relativistic corrections cancel each other in these decays.

\end{abstract}
\maketitle
\section{Introduction}
As the complement of ground-state particles, the study of excited-state particles has attracted increasing attention, as they contain richer physical content, and their study is more complicated than that of ground-state particles. For example, the relativistic correction of an excited-state particle is larger than that of its ground state, which needs to be treated carefully \cite{rela}.

The $B^{*}_{2}$(5747)$^{0}$ ($B^{*}_{2}$(5747)$^{\pm}$) is the orbitally excited state of the pseudoscalar $B$ meson, and it is also the $2^+$ ground state. $B^{*}_{2}$(5747)$^{0}$ was discovered by D0 Collaborations \cite{find1} in 2007; by its decaying to $B^{(*)+}\pi^-$, its mass is detected as $M(B_2^*)=5746.8\pm2.4\pm1.7$ MeV. Later, this state was confirmed by CDF Collaboration \cite{find2,find3}.
In 2015, the LHCb collaboration\cite{jhep1504} made precise measurements of the mass and decay width of the $B^{*}_{2}$(5747)$^{0}$. Now, the average mass and decay width of $B^{*}_{2}$(5747)$^{0}$ listed in PDG \cite{find4} are $M(B_2^*)=5739.5\pm0.7$ MeV and $\Gamma(B_2^{*0})=24.2\pm1.7$ MeV,respectively.

The dominant decay channels of $B^{*}_{2}$(5747)$^{0}$ are the OZI-allowed strong decays, $B^{*}_{2}$(5747)$^{0}$ $\rightarrow$ $B^{+}$ $\pi^{-}$, $B^{*+}$ $\pi^{-}$, $B^{0}$ $\pi^{0}$ and $B^{*0}$ $\pi^{0}$. Other decay channels, like the electromagnetic decays, can be ignored compared to the OZI-allowed strong decays. The OZI-allowed strong decays have been studied by various theoretical models \cite{falk,zsl,orsland,pierro,wenzhang1,wenzhang2,wangzg2,wenzhang3,wenzhang4,
wenzhang5,gupta,wangzg,zhongxh},
but there are still some interesting questions to be studied on this topic. For example, the relativistic corrections of this process are not well studied, especially those introduced by heavy $B^{*}_{2}$(5747)$^{0}$ and $B^{(*)}$ mesons. We have found that even in a process where double heavy mesons are involved, the relativistic corrections are large and important \cite{gengzikan}, so the relativistic correction in a heavy$-$light meson is expected to be large.

Therefore, in this paper, we will study the strong decays of $B^{*}_{2}$(5747)$^{0}$ by using the instantaneous Bethe$-$Salpeter method, emphasis is paid on the relativistic corrections. The Bethe$-$Salpeter equation \cite{find5} is the relativistic dynamic equation describing two-body bound state; its instantaneous version is called the Salpeter equation  \cite{find6}. We will solve the Salpeter equation to obtain the numerical wave functions used in this paper. The partial waves of the corresponding wave functions will be given and discussed, since they are important to provide the relativistic corrections.

In Sect.2, we will show the method to calculate the transition matrix element.The wave functions used and their partial waves, as well as the fraction ratios of partial waves, are given in Sect.3. The theoretical decay width and the contributions of different partial waves are given in Sect.4, we give a short summary in Sect.5.

\section{Hadronic transition matrix elements}
In this section, we take the decay channel $B^{*}_{2}$(5747)$^{0}$ $\rightarrow$ $B^{+}$$\pi^{-}$ as an example to show our method to calculate the transition amplitude. Since we will solve the instantaneous Bethe$-$Salpeter equation to obtain the wave functions of mesons, but instantaneous approximation is only good for a heavy meson, and not for a light meson, the wave function of $\pi$ may bring us large errors, so we abandon choosing the widely used $^3P_0$ model, where the $\pi$ meson wave function is needed.

In our method, we adopt the reduction formula to avoid using the wave function of $\pi$. After further using the partial conservation of the axial vector current (PCAC) and low energy theory, the transition $S$-matrix for the decay $B^{*}_{2}$(5747)$^{0}$ $\rightarrow$ $B^{+}$$\pi^{-}$ can be written as \cite{find14}
\begin{eqnarray}
\langle B^{+}(P_{f})\pi^{-}(P_{\pi})|B^{*0}_{2}(P)\rangle\nonumber\\&\quad= (2\pi)^4\delta^4(P-P_f-P_{\pi})\frac{iP^{\mu}_{\pi}}{f_{\pi}}\nonumber\\&\qquad\times\langle B^{+}(P_{f})|\bar{d}\gamma_{\mu}\gamma_{5}{u}|B^{*0}_{2}(P)\rangle,
\end{eqnarray}
where $P$, $P_{f}$, and $P_{\pi}$ are momenta of ${B}^{*0}_{2}$, $B^{+}$, and $\pi^{-}$, respectively, and $f_{\pi}$ is the decay constant of $\pi^{-}$.
Then the transition amplitude for the process $B^{*}_{2}(5747)\rightarrow B^{+}\pi^-$ is
\begin{eqnarray}
\mathcal{M}=\frac{iP_{\pi}^{\mu}}{f_{\pi}}\langle B^{+}(P_{f})|\bar d\gamma_{\mu}\gamma_{5}u|B^{*0}_{2}(P)\rangle.
\end{eqnarray}

By adopting the Mandelstam formalism \cite{find13}, the hadron matrix element can be written as \cite{find14}
\begin{eqnarray}\label{transition}
&&\langle B^{+}(P_{f})|\bar d\gamma_{\mu}\gamma_{5}{u}|B^{*0}_{2}(P)\rangle=
\int\frac{d^3q_{_\bot}}{(2\pi)^{3}}Tr\left[\bar{\varphi}^{++}_{P_f}(q_{_{f_{\bot}}})
\gamma_{\mu}\gamma_{5}\varphi^{++}_{P}(q_{_\bot})\frac{\slashed{P}}{M}\right]
\end{eqnarray}
where $q_{_\bot}$ is the instantaneous relative momentum between the quark and antiquark in the initial meson, and ${\varphi}_{P_f}^{++}(q_{_{f_{\bot}}})$ and $\varphi_{P}^{++}(q_{_\bot})$ are the positive-energy wave functions of $B^{+}$ and $B^{*0}_{2}$, respectively. We will show the details of them in the next section.

\section{Wave function and its partial waves}

\subsection{$2^+$ state}
For a $2^+$ state $B^{*0}_{2}$, its relativistic wave function can be written as \cite{wenzhang8}
\begin{eqnarray}\label{2+ wave}
\varphi_{2^{+}}(q_{_\bot})=&\varepsilon_{\mu\nu}^{(\lambda)}q_{\bot}^{\nu}\left
[q_{\bot}^{\mu}\left(f_{1}+\frac{\slashed{P}}{M}f_{2}+\frac{\slashed{q}_{\bot}}{M}f_{3}+
\frac{\slashed{q}_{\bot}\slashed{P}}{M^{2}}f_{4}\right)+\gamma^{\mu}\right.\nonumber\\
&\left.\times\left( Mf_{5}+\slashed{P}f_{6}+
\slashed{q}_{\perp}f_{7}+\frac{i}{M}f_{8}\epsilon_{\mu\alpha\beta\gamma}P^{\alpha}
q_{\perp}^{\beta}\gamma^{\gamma}\gamma_{5}\right)\right],\nonumber\\
\end{eqnarray}
where $\varepsilon_{\mu\nu}^{(\lambda)}$ is the polarization tensor of the $2^{+}$ state; $M$ and $P$ are the mass and momentum, respectively; ${q}_{_{\bot}}$ is the three-dimensional relative momentum between the quark and antiquark with the definition ${q}_{_{\bot}}=q-\frac{P\cdot q}{M^2}P$, in the center-of-mass frame of the state ${q}_{_{\bot}}=(0,\vec{q})$; and the radial wave function $f_i$ ($i=1,2,\ldots,8$) is a function of $\vec{q}^2=-q^2_{_\bot}$.
Similar to the $S-D$ mixing in the $1^-$ state $\Psi(3770)$, there is possible $P-F$ mixing in a $2^+$ state. But in our relativistic method, we pointed out that \cite{bc wave}instead of $P-F$ mixing, we have $P-D-F$ mixing in a $2^+$ state. In Eq.(\ref{2+ wave}), the terms including $f_5$ and $f_6$ are $P$ waves, the terms with $f_3$ and $f_4$ are $F$ waves, and the others are all $D$ waves. For the ground $2^+$ state $B^{*}_2(5747)$, in a non-relativistic limit, it is a pure $P$ wave, so the $D$ and $F$ partial waves will provide the relativistic corrections. Following the method in Ref. \cite{bc wave}and considering the normalization condition, we calculate the case of $B^{*}_2(5747)$ and obtain the ratios
\begin{equation}P/D/F=1:0.421:0.051.\end{equation}
Therefore, as the ground $2^+$ state, its wave function is $P$ wave dominant, but with a large amount of the $D$ wave and a small amount of the $F$ wave. In a previous paper \cite{bc wave}, we obtained $P/D/F=1:0.10:0.039$ for $B^{*}_{c2}(1P)$. So the content of the $D$ wave in the $B^{*}_2(5747)$ meson is much larger than that in the $B^{*}_{c2}(1P)$ meson, which means that the relativistic correction of the former is much larger than that of the latter.
The upper results of $P$, $D$, and $F$ components are obtained by only considering the wave function and its normalization, and when we study a transition process, the conclusion may be changed. This is because first, it is the positive wave function, not the wave function itself, taking part in a hadronic transition, see Eq.(\ref{transition}); second, in a particular process, different partial waves will behave differently and contribute differently. For example, $\Psi(3770)$ includes $S$, $P$, and $D$ waves, but in its dilepton annihilation decay, only the $S$ wave has a contribution. Thus we wish to study the contributions of different partial waves in the strong decays of $B^{*}_2(5747)$, but we will show the results in the next section. Here we only show the positive part of the $2^+$ wave function Eq.(\ref{2+ wave}),
\begin{eqnarray}\label{2++}
\varphi_{2^{+}}^{++}(q_{_\bot})=\varepsilon_{\mu\nu}^{(\lambda)}q_{\bot}^{\nu}
\left[q_{\bot}^{\mu}\left(A_{1}+A_{2}\frac{\slashed{P}}{M}+A_{3}\frac{\slashed{q}_{\bot}}{M}+
A_{4}\frac{\slashed{q}_{\bot}\slashed{P}}{M^{2}}\right)\right.\nonumber\\
+\left.\gamma^{\mu}\left(A_{5}+A_{6}\frac{\slashed{P}}{M}+A_{7}\frac{\slashed{q}_{\bot}}{M}
+A_{8}\frac{\slashed{P}\slashed{q}_{\bot}}{M^{2}}\right)\right],\nonumber\\
\end{eqnarray}
where the expressions of $A_i$ $(i=1,2,\ldots,8)$ are shown in the Appendix.
Similar to the original one, the $A_5$ and $A_6$ terms are $P$ waves, $A_3$ and $A_4$ terms are $F$ waves, and the others are $D$ waves.
\subsection{$1^-$ state}
The final meson could be a $1^-$ vector $B^*$, whose wave function is \cite{1-}
\begin{eqnarray}
\varphi_{1^{-}}(q_{_{f\bot}})=&\epsilon_{_f}\cdot q_{_{f\bot}}\left(g_{1}+\frac{\slashed{P_f}}{M_{f}}g_{2}+\frac{\slashed{q}_{f\bot}}{M_{f}}g_{3}+
\frac{\slashed{P_f}\slashed{q}_{f\bot}}{M_{f}^{2}}g_{4}\right)+M_{f}\slashed{\varepsilon}_{_f}g_{5}
+\slashed{\varepsilon}_{_f}\slashed{P_f}g_{6}\nonumber\\
&+\left(\slashed{q}_{f\bot}\slashed{\varepsilon}_{_f}-\epsilon_{_f}\cdot q_{_{f\bot}}\right)g_{7}+
\frac{1}{M_{f}}\left(\slashed{P_f}\slashed{\varepsilon}_{_f}\slashed{q}_{f\bot}-\slashed{P_f}\epsilon_{_f}\cdot q_{_{f\bot}}\right)g_{8},
\end{eqnarray}
where the $\epsilon^{\mu}_{_f}$ is the polarization vector of the $B^*$ meson. This wave function includes $S$ partial waves ($g_5$ and $g_6$ terms), $P$ partial waves ($g_1$, $g_2$, $g_7$, and $g_8$ terms), and $D$ partial waves ($g_3$ and $g_4$ terms). So the $B^*$ meson is a $S-P-D$ mixing state; in the non-relativistic limit, it is a pure $S$ wave state, and $P$ and $D$ partial waves are relativistic corrections. For $B^*$, we obtain \begin{equation}S/P/D=1:0.354:0.046.\end{equation}
This result shows us that the relativistic correction of the $B^*$ meson is much larger than that of $B_c^*$, since for the latter, we have $S/P/D=1:0.09:0.037$ \cite{bc wave}.
The positive part of the $B^*$ wave function can be written as
\begin{eqnarray}\label{1--}
\varphi_{1^{-}}^{++}(q_{_{f\bot}})=&\epsilon_{_f}\cdot q_{_{f\bot}}\left(B_{1}+B_{2}\frac{\slashed{P_{f}}}{M_{f}}+B_{3}\frac{\slashed{q}_{f\bot}}{M_{f}}
+B_{4}\frac{\slashed{q}_{f\bot}\slashed{P_f}}{M_{f}^{2}}\right)\nonumber\\
&+M_{f}\slashed{\epsilon}_{_f}\left(B_{5}+B_{6}\frac{\slashed{P_f}}{M_{f}}
+B_{7}\frac{\slashed{q}_{f\bot}}{M_{f}}+B_{8}\frac{\slashed{q}_{f\bot}\slashed{P_f}}{M_{f}^{2}}\right),
\end{eqnarray}
and we give the detailed expressions of $B_is$ in the Appendix. Similarly, $B_5$ and $B_6$ terms are $S$ waves, $B_3$ and $B_4$ terms are $D$ waves, and the others are $P$ waves.
\subsection{$0^-$ state}
Another final meson is the pseudoscalar $B$ meson, which is a $0^-$ state, and its wave function can be written as \cite{0-}
\begin{equation}
\varphi_{0^{-}}(q_{_{f\bot}})=\left(\slashed{P_f}h_{1}+M_{f}h_{2}
+\slashed{q}_{f\perp}h_{3}+\frac{\slashed{P_f}\slashed{q}_{f\perp}}{M_{f}}h_{4}\right)\gamma_{5},
\end{equation}
where $h_1$ and $h_2$ terms are $S$ waves, while $h_3$ and $h_4$ terms are $P$ waves. We obtain the following result for the $B$ meson,
\begin{equation}S/P=1:0.343,\end{equation}
which is very different from the $B_c$ case, $S/P=1:0.082$, showing that the relativistic correction in the $B$ meson is much larger than that in the $B_c$ meson.
We also note that if we ignore the tiny $D$ wave in $B^*$ meson, we have the relation $0.354\approx0.343$ for the $P$ wave components in $B$ and $B^*$, which means that the relativistic corrections for them are approximately equal.
The positive wave function for $B$ is
\begin{equation}\label{0-+}
\varphi_{0^{-}}^{++}(q_{_{f\bot}})=\left(C_{1}+C_{2}\frac{\slashed{P_f}}{M_{f}}
+C_{3}\slashed{q}_{f\bot}+C_{4}\frac{\slashed{q}_{f\bot}\slashed{P_f}}{M_{f}}\right)\gamma_{5},
\end{equation}
where the expressions of $S$ partial waves, $C_1$ and $C_2$, and $P$ partial waves, $C_3$ and $C_4$, are given in the Appendix.
\section{Strong decay widths and contributions of partial waves}
\subsection{Strong decay widths}
In the numerical calculation, the following parameters are used:
$m_{b}=4.96$ GeV, $m_{u} = 0.305$ GeV, $m_{d}=0.311$ GeV, and $ f_{\pi} = 0.1307$ GeV. Our relativistic results of dominant strong decays of $B^{*}_{2}$(5747)$^{0}$ are
\begin{equation}
\Gamma(B^{+}\pi^-)=9.05~ {\rm MeV,}\quad\Gamma(B^{*+}\pi^-) = 8.47 ~ {\rm MeV,}\nonumber
\end{equation}
\begin{equation}
\Gamma(B^{0}\pi^0) = 4.19 ~ {\rm MeV,}\quad\Gamma(B^{*0}\pi^0) = 4.23 ~ {\rm MeV.}
\end{equation}
The estimated total width is 25.9 MeV. The following relative branching fractions are also calculated.
\begin{equation}
R_0=\frac{\Gamma(B^{*}\pi)}{\Gamma(B\pi)} = \frac{\Gamma(B^{*+}\pi^-)+\Gamma(B^{*0}\pi^0)}{\Gamma(B^{+}\pi^-)+\Gamma(B^{0}\pi^0)} = 0.96,\nonumber
\end{equation}
\begin{equation}
R_1=\frac{\Gamma(B^{*+}\pi^-)}{\Gamma(B^{+}\pi^-)} = 0.94,\quad R_2 = \frac{\Gamma(B^{*0}\pi^0)}{\Gamma(B^{0}\pi^0)} = 1.0.
\end{equation}
To compare with other results, we show our predictions and other theoretical results as well as experimental data in Table \ref{tab1}. We can see that our predictions of widths are very consistent with the experimental data, and also agree well with the theoretical results in Refs. \cite{wenzhang3,wangzg} For the relative branching fraction, all the theoretical results agree with experimental data within an acceptable error range.
\begin{table}[ht]
\centering \caption{Decay width of $B^{*}_{2}$(5747)$^{0}$, in units of MeV} \label{tab1}
\setlength{\tabcolsep}{6pt} 
\renewcommand{\arraystretch}{1} 
\begin{tabular}{|c|c|c|c|c|c|c|c|c|c|}
\hline
     & Ours & \cite{orsland} & \cite{wenzhang1}      & \cite{wenzhang3}   & \cite{wenzhang4}    & \cite{wenzhang5} &\cite{wangzg}& \cite{zhongxh} & $\Gamma_{exp}$\cite{find4} \\ \hline
             $B\pi$   & 13.2 & 14.7& 25  & 12.62 & 6.23  & 9.77  &13.3 &16.3&                    \\ \hline
        $B^{*}\pi$ & 12.7&13.8 & 22 & 11.89 & 5.04  & 9.79  & 11.4 &15.4&    \\\hline
              Total      & 25.9& 28.5 & 47   & 24.51 & 11.27 & 19.56 &24.7&32& 24.2$\pm$1.7            \\ \hline
      $\frac{\Gamma(B^*\pi)}{\Gamma(B\pi)}$  & 0.96 & 0.94& 0.88   &0.94   & 0.81 & 1.0  &0.86& 0.94& 0.84$\pm$0.27    \\\hline 
\end{tabular}\\
\centering \caption{Decay widths (MeV) of different partial waves, where ``whole" means the complete wave function of the $2^+$ $B^{*}_{2}(5747)^{0}$ or $0^-$ $B^{*}$ meson.}
\label{tab2}
\setlength{\tabcolsep}{6pt}
\renewcommand{\arraystretch}{1}
\begin{tabular}{|c|c|c|c|}
\hline
{\diagbox{$2^+$}{$0^-$}}& whole & S wave ($C_1$,$C_2$)& P wave ($C_3$,$C_4$) \\ \hline
  whole   & 9.05 & 7.21 & 0.074 \\ \hline
  P wave ($A_5$,$A_6$) & 15.6 &10.4 &0.518  \\ \hline
  D wave ($A_1$,$A_2$,$A_7$,$A_8$)& 0.969& 0.291& 0.198 \\ \hline
  F wave ($A_3$,$A_4$) & $\sim$ 0 &$\sim$ 0 & $\sim$ 0 \\ \hline
\end{tabular}\\
\centering \caption{Decay widths (MeV) of different partial waves, where ``whole" means the complete wave function of $2^+$ $B^{*}_{2}(5747)^{0}$ or $1^-$ $B^*$ meson.}
\label{tab3}
\setlength{\tabcolsep}{6pt}
\renewcommand{\arraystretch}{1}
\begin{tabular}{|c|c|c|c|c|}
\hline
{\diagbox{$2^+$}{$1^-$}}&whole&S wave ($B_5$,$B_6$) &P wave ($B_1$,$B_2$,$B_7$,$B_8$)&D wave ($B_3$,$B_4$) \\ \hline
 whole   & 8.47 & 6.82 &0.091 & $\sim$ 0\\ \hline
  P wave ($A_5$,$A_6$) & 14.6&9.37 &0.592  &$\sim$ 0\\ \hline
  D wave ($A_1$,$A_2$,$A_7$,$A_8$)&0.819&0.201 &0.209 & $\sim$ 0\\ \hline
  F wave ($A_3$,$A_4$) & $\sim$ 0 &$\sim$ 0 &$\sim$ 0  &0\\ \hline
\end{tabular}
\end{table}

\clearpage 
Takeing the processes $B^{*}_{2}(5747)^{0}\rightarrow B^{+}\pi^{-}$ and $B^{*}_{2}(5747)^{0}\rightarrow B^{*+}\pi^{-}$ as examples,
we study the contributions of different partial waves in initial and final mesons, the results are shown in Table \ref{tab2} and Table \ref{tab3}, respectively. In the tables, `$2^+$' means the initial meson $B^{*}_{2}(5747)^{0}$ is a $J^P=2^+$ state, its complete wave function is denoted by `$whole$', which includes $P$ ($A_5$ and $A_6$ terms in Eq.(\ref{2++})), $D$ ($A_1$,$A_2$,$A_7$ and $A_8$ terms) and $F$ ($A_3$ and $A_4$ terms) partial waves; `$0^-$' and `$1^-$' stand for the final heavy pseudoscalar $B$ (whose wave function includes $S$ and $P$ partial waves) and vector $B^*$ (whose wave function contains $S$, $P$ and $D$ partial waves), respectively.
Table \ref{tab2} shows us that when the wave functions of the initial $2^+$ state $B^{*}_{2}(5747)^{0}$ and final $0^-$ $B$ meson are all completely relativistic, the decay width is $\Gamma(B^{+}\pi^-)=9.05$ MeV; in the non-relativistic limit, the wave function of $B^{*}_{2}(5747)^{0}$ is a pure $P$ wave, only includes $A_5$ and $A_6$ terms in Eq.(\ref{2++}), while the wave function of $B$ is a pure $S$ wave and only has $C_1$ and $C_2$ terms in Eq.(\ref{0-+}). Then the decay width in a non-relativistic limit without relativistic correction is
\begin{equation}
\Gamma_{0}(B^{+}\pi^-)=10.4~ \rm{MeV},
\end{equation}
a little larger than the relativistic result. The relativistic effect can be defined as
\begin{equation}
\frac{\Gamma_{0}(B^{+}\pi^-)-\Gamma(B^{+}\pi^-)}{\Gamma(B^{+}\pi^-)}=0.15.
\end{equation}
This value is much smaller than our expected VALUE, since in the $2^+$ wave function we have $P:D=1:0.42$, and we have $S:P=1:0.34$ in $0^-$ wave function. These ratios are consistent with the following result; when the wave function of $B$ meson is complete, while the wave function of $B^{*}_{2}(5747)^{0}$ only contains the non-relativistic $A_5$ and $A_6$ terms, the decay width is $15.6$ MeV, which is our expectation, because it is much larger than the relativistic $9.05$ MeV. The obtained small relativistic effect means that the relativistic corrections cancel each other in this process. We also show other possible overlaps of initial and final partial waves in Table \ref{tab2}; for example, the contribution of the $F$ partial wave in the $2^+$ state is very small, $\sim 0$ (several orders smaller), and can be safely ignored .
Table \ref{tab3} shows us that, when the wave functions of the $B^{*}_{2}(5747)^{0}$ and final $B^*$ meson are all complete, the decay width is $\Gamma(B^{*+}\pi^-)=8.47$ MeV; in the non-relativistic limit, besides the $B^{*}_{2}(5747)^{0}$ being a pure $P$ wave, the $B^*$ is a pure $S$ wave whose wave function only contains $B_5$ and $B_6$ terms in Eq.(\ref{1--}). Then the decay width in the non-relativistic limit is
\begin{equation}
\Gamma_{0}(B^{*+}\pi^-)=9.37~ \rm{MeV}.
\end{equation}
The relativistic effect is
\begin{equation}
\frac{\Gamma_{0}(B^{*+}\pi^-)-\Gamma(B^{*+}\pi^-)}{\Gamma(B^{*+}\pi^-)}=0.11,
\end{equation}
which is also very small but comparable to the upper case of the $B$ meson.
When the final wave function of the $B^*$ meson is completely relativistic while the initial one is non-relativistic, the decay width is about $14.6$ MeV, much larger than the relativistic $8.47$ MeV. This indicates that the relativistic correction is very large in the $B^{*}_{2}(5747)^{0}$ meson, which is consistent with the large $D$ wave component in the wave function since $P:D=1:0.421$. The contributions of the $F$ partial wave for the $B^{*}_{2}(5747)^{0}$ meson and $D$ wave for $B^*$ are ignorable since their results ($\sim0$) are several orders smaller than other partial waves. We also note that in the transition of $B^{*}_{2}(5747)^{0}\to B^*\pi$, there is no overlap of the $F$ wave in the $B^{*}_{2}(5747)^{0}$ meson and the $D$ wave in the $B^*$ meson.

\section{Summary}
In this paper, we present a relativistic study on the two-body OZI-allowed strong decays of the $B^{*}_{2}$(5747)$^{0}$ meson by using the Bethe$-$Salpeter method. The estimated full width $\Gamma(B^{*}_{2}$(5747)$^{0})=25.9$ MeV and the branching fraction $\frac{\Gamma(B^*\pi)}{\Gamma(B\pi)}=0.96$ are highly consistent with experimental data.

The contributions of different partial waves in the initial and final mesons are also studied. First, we find that the wave function of $B^{*}_{2}$(5747)$^{0}$ is $P$ wave-dominant, but with a large amount of $D$ wave and a small amount of $F$ wave. The large $D$ wave component will lead to a large relativistic correction, which is confirmed by the large difference between the complete result and that where only the $P$ wave is considered in the $B^{*}_{2}$(5747)$^{0}$ meson. Similarly, the relativistic corrections are large for the heavy$-$light pseudoscalar $B$ and vector $B^*$, since we find that they are an $S$ wave-dominant state but contain a large amount of $P$ wave. Third, the large relativistic terms in the wave functions of the initial and final mesons do not lead to a large corrections in the transition of $B^{*}_{2}$(5747)$^{0}\to B\pi$ and $B^{*}_{2}$(5747)$^{0}\to B^*\pi$, showing us that there are cancellations between the relativistic corrections.

\vspace{0.7cm} {\bf Acknowledgments}

This work was supported in part by the National Natural Science Foundation of China (NSFC) under Grant No. 12075073, the Natural Science Foundation of Hebei province under Grant No. A2021201009, and the Post-graduate's Innovation Fund Project of Hebei University under Grant No. HBU2022BS002. Data Availability This manuscript has no associated data or the data will not be deposited.
\appendix
\section{Appendix:Bethe$-$Salpeter wave function}
In this section we give the representation of the coefficients of the positive energy wave functions of the $2^{+}$, $1^{-}$, and $0^{-}$ states. For the $2^{+}$ state,
\begin{eqnarray}
&&J\equiv f_{3}+f_{4}\frac{m_{1}+m_{2}}{\omega_{1}+\omega_{2}},  \quad
K\equiv f_{5}-f_{6}\frac{\omega_{1}+\omega_{2}}{m_{1}+m_{2}},\nonumber\\
&&A_{1}=\frac{(\omega_{1}+\omega_{2})q_{\bot}^{2}}{2M(m_{1}\omega_{2}
+m_{2}\omega_{1})}J+\frac{(f_{5}\omega_{2}-
f_{6}m_{2})M}{m_{1}\omega_{2}+m_{2}\omega_{1}},\nonumber\\
&&A_{2}=\frac{(m_{1}+m_{2})q_{\bot}^{2}}{2M(m_{1}\omega_{2}+m_{2}\omega_{1})}J
+\frac{(f_{6}\omega_{2}-f_{5}m_{2})M}{m_{1}\omega_{2}+m_{2}\omega_{1}},\nonumber\\
&&A_{3}=\frac{1}{2}J-\frac{f_{6}M^{2}}{m_{1}\omega_{2}+m_{2}\omega_{1}},\quad
A_{4}=-\frac{1}{2}\frac{\omega_{1}+\omega_{2}}{m_{1}+m_{2}}J
+\frac{f_{5}M^{2}}{m_{1}\omega_{2}+m_{2}\omega_{1}},\nonumber\\
&&A_{5}=\frac{M}{2}K,\quad\quad\quad\quad\quad\quad\quad
A_{6}=-\frac{M(m_{1}+m_{2})}{2(\omega_{1}+\omega_{2})}K,\nonumber\\
&&A_{7}=\frac{M^{2}(\omega_{1}-\omega_{2})}{2(m_{2}\omega_{1}+m_{1}\omega_{2})}K,\quad
A_{8}=-\frac{M^{2}(m_{1}+m_{2})}{2(m_{2}\omega_{1}+m_{1}\omega_{2})}K,\nonumber
\end{eqnarray}
where $\omega_{i}=\sqrt{m_i^2-q_{\bot}^{2}}$ with $i=1,2$.

For the $1^{-}$ state, we have
\begin{eqnarray}
&&B_{1}=\frac{-(\omega_{1f}+\omega_{2f})q_{_{f_{\bot}}}^{2}g_{3}-(m_{1f}
+m_{2f})q_{_{f_{\bot}}}^{2}g_{4}+2M_{f}^{2}\omega_{2f}g_{5}-2M_{f}^{2}m_{2f}g_{6}}{2M_{f}(m_{1f}\omega_{2f}+m_{2f}\omega_{1f})},\nonumber\\
&&B_{2}=\frac{-(m_{1f}+m_{2f})q_{_{f_{\bot}}}^{2}g_{3}-(\omega_{1f}
-\omega_{2f})q_{_{f_{\bot}}}^{2}g_{4}+2M_{f}^{2}\omega_{2f}g_{6}-2M_{f}^{2}m_{2f}g_{5}}{2M_{f}(m_{1f}\omega_{2f}+m_{2f}\omega_{1f})},\nonumber\\
&&B_{3}=\frac{1}{2}\left(g_{3}+\frac{m_{1f}+m_{2f}}{\omega_{1f}+\omega_{2f}}g_{4}
-\frac{2M_{f}^{2}}{m_{1f}\omega_{2f}+m_{2f}\omega}_{1}g_{6}\right), \nonumber\\
&&B_{4}=\frac{1}{2}\left(g_{4}+\frac{\omega_{1f}+\omega_{2f}}{m_{1f}+m_{2f}}g_{3}
-\frac{2M_{f}^{2}}{m_{1f}\omega_{2f}+m_{2f}\omega}_{1}g_{5}\right), \nonumber\\
&&B_{5}=\frac{1}{2}\left(g_{5}-\frac{\omega_{1f}+\omega_{2f}}{m_{1f}+m_{2f}}g_{6}\right),\quad
B_{6}=\frac{1}{2}\left(g_{6}-\frac{m_{1f}+m_{2f}}{\omega_{1f}+\omega_{2f}}g_{5}\right),\nonumber\\
&&B_{7}=\frac{M_{f}}{2}\frac{\omega_{1f}-\omega_{2f}}{m_{1f}\omega_{2f}
+m_{2f}\omega_{1f}}\left(g_{5}-\frac{\omega_{1f}+\omega_{2f}}{m_{1f}+m_{2f}}g_{6}\right),\nonumber\\
&&B_{8}=\frac{M_{f}}{2}\frac{m_{1f}+m_{2f}}{m_{1f}\omega_{2f}+m_{2f}\omega_{1f}}
\left(-g_{5}+\frac{\omega_{1f}+\omega_{2f}}{m_{1f}+m_{2f}}g_{6}\right).\nonumber
\end{eqnarray}

For the $0^{-}$ state, we have,
\begin{eqnarray}
&&C_{1}=\frac{M_{f}}{2}\left(h_{1}\frac{\omega_{1f}+\omega_{2f}}{m_{1f}+m_{2f}}+h_{2}\right),
\quad C_{3}=-C_2\frac{(m_{1f}-m_{2f})}{m_{1f}\omega_{2f}
+m_{2f}\omega_{1f}},\nonumber\\
&&C_{2}=\frac{M_{f}}{2}\left(h_{1}+h_{2}\frac{m_{1f}+m_{2f}}{\omega_{1f}+\omega_{2f}}\right),
\quad C_{4}=C_1\frac{\omega_{1f}+\omega_{2f}}{m_{1f}\omega_{2f}
+m_{2f}\omega_{1f}}.\nonumber\\
\nonumber
\end{eqnarray}

\end{document}